\documentclass[%
 reprint,
 superscriptaddress,
 showpacs,preprintnumbers,
 amsmath,amssymb,
 aps,
 prl,
 longbibliography,
lengthcheck,%
]{revtex4-1}

\usepackage{graphicx}
\usepackage{dcolumn}
\usepackage{bm}
\usepackage{color}
\usepackage{ulem}
\usepackage{hyperref}

\begin{document}

\preprint{APS/123-QED}

\title{
Spin motive force by the momentum-space Berry phase in magnetic Weyl semimetals
}

\author{Akira Harada}
\affiliation{
Department of Physics, Tokyo Institute of Technology, Meguro, Tokyo, 152-8551, Japan
}

\author{Hiroaki Ishizuka}
\affiliation{
Department of Physics, Tokyo Institute of Technology, Meguro, Tokyo, 152-8551, Japan
}

\date{\today}

\begin{abstract}
We show that the magnetic precession of ferromagnetic moments in a noncentrosymmetric magnetic Weyl semimetal induces an electric current through a mechanism analogous to the adiabatic charge pumping.
The current is a consequence of a Berry phase in the momentum space resulting from the circular motion of Weyl nodes induced by the precession.
This mechanism resembles the Faraday effect, namely, induced magnetic field by circular electric current.
The circular motion of Weyl nodes induces magnetic charge current in the momentum space, which results in a Berry phase that describes the adiabatic pump.
Experimentally, this phenomenon is similar to spin motive force, which is an electric current induced by magnetic precision in the presence of the spatial gradient of magnetization.
However, unlike the conventional spin motive force, this current occurs without a magnetization gradient.
The result demonstrates a nontrivial interplay between the topological electronic state and magnetic dynamics.
\end{abstract}

\pacs{
}

\maketitle


{\it Introduction} --- The interplay of topological electronic states and magnetism realize peculiar electronic states and phenomena arising from the interplay, which attract attention from both basic science and application such as spintronics~\cite{He2022}.
In topological insulators~\cite{Hasan2010,Qi2011}, introducing ferromagnetism makes it a quantum anomalous Hall insulator~\cite{Chang2013,Checkelsky2014,Kou2014}, and coupling topological insulators to antiferromagnets were discussed as a route to realize axion insulator~\cite{Qi2008}.
Effect of magnetism on non-trivial electronic states were also studied in relation to Weyl semimetals (WSM)~\cite{Murakami2007,Burkov2011,Wan2011,Yan2017,Armitage2018} such as in pyrochlore iridates~\cite{Wan2011,Krempa2012,Krempa2014}.
In these materials, their transport properties were discussed in relation to Weyl electrons, such as relatively large Hall effect with a small magnetization~\cite{Moon2013,Ueda2018,Shekhar2018}, and non-monotonic magnetic-field dependence of anomalous Hall effect in EuTiO$_3$~\cite{Takahashi2018}.
Understanding the effect of coupling between magnetic order and the topological electronic states was key to understanding the properties of these properties.

On the other hand, dynamical or optical properties of magnetic materials bring about rich functionalities~\cite{Maekawa2017}, the study of which has been one of the major topics in spintronics.
Related studies on Weyl semimetals find novel phenomena related to the axial field arising from domain walls~\cite{Araki2018,Hannukainen2020,Hannukainen2021}.
While many studies were done for the effect of magnetic textures, much less is known about the role of the topological/geometrical nature of electrons on the dynamical properties.
In this work, we study the electrical response of WSM due to the magnetic dynamics as an illustration of the interplay, in which we find an electrical current analogous to adiabatic charge pumping~\cite{Thouless1983,Niu1984} induced by the precession of magnetic moment.

A quantity that plays the central role in this study is the Berry phase of electronic bands defined by $\bm b_{\bm kn}=\nabla_k\times\bm a_{\bm kn}$, where $\nabla_k=(\partial_{kx},\partial_{ky},\partial_{kz})$ and $\bm a_{\bm kn}=-{\rm i}\left<u_{n\bm k}\right|\nabla_k\left|u_{n\bm k}\right>$ is the berry connection with $\left|u_{n\bm k}\right>$ being the Bloch function of $n$th band with momentum $\bm k$~\cite{Sandaram1999,Xiao2010}.
The Berry curvature is directly related to the topological nature of electronic states such as the Hall conductivity in quantum~\cite{Thouless1982} and anomalous Hall effects~\cite{Karplus1954}.

Another system with a characteristic feature in $\bm b_{\bm kn}$ is WSM, where the Weyl node is a singular point of the Berry curvature with $\bm b_{\bm kn}\propto (\bm k-\bm k_0)/|\bm k-\bm k_0|^3$ ($\bm k_0$ is the position of the Weyl node).
The distribution of $\bm b_{\bm kn}$ resembles that of the magnetic field around a point magnetic charge, hence sometimes called magnetic monopole in the momentum space.
The divergent Berry curvature at the Weyl node is often related to the unique properties of WSM.
For instance, it gives rise to the non-monotonic magnetization dependence of anomalous Hall conductance~\cite{Takahashi2018} and enhances the electromagnetic response related to Berry curvature as in anomaly-related magnetoresistance~\cite{Son2013}.
The latter is generally possible in a system with a finite Berry curvature, but the large Berry curvature around the Weyl nodes enhances the phenomenon~\cite{Ishizuka2019a,Ishizuka2019b}.
The enhanced Berry phase effect make WSM a particularly interesting material for studying the Berry-phase-related phenomena.

Beside the Berry curvature in the momentum space $\bm b_{\bm kn}$, a Berry curvature with time derivative
\begin{align}
    \bm e_{\bm kn}(t)=\partial_t\bm a_{\bm kn}(t)-\nabla_k a^t_{\bm kn}(t)
\end{align}
also contributes to electron transport in a system where the Hamiltonian $H(t)$ evolves adiabatically over time, a phenomenon known as adiabatic pump~\cite{Thouless1983,Niu1984}.
Here, $\bm a_{\bm kn}(t)=-{\rm i}\left<u_{n\bm k}(t)\right|\nabla_k\left|u_{n\bm k}(t)\right>$ is the Berry connection defined by the Bloch function for the instantaneous Hamiltonian $H(t)$ and $a^t_{\bm kn}(t)=-{\rm i}\left<u_{n\bm k}(t)\right|\partial_t\left|u_{n\bm k}(t)\right>$.
However, the charge pump requires a time-dependent perturbation comparable to the bandwidth, and hence, it is realized only in cold atoms~\cite{Nakajima2016,Lohse2016}.
An exception to the obstruction might be WSM, in which the divergent Berry curvature near the Weyl nodes enhances the $\bm e_{\bm kn}(t)$ field~\cite{Ishizuka2016,Ishizuka2017b}.

\begin{figure}
  \includegraphics[width=\linewidth]{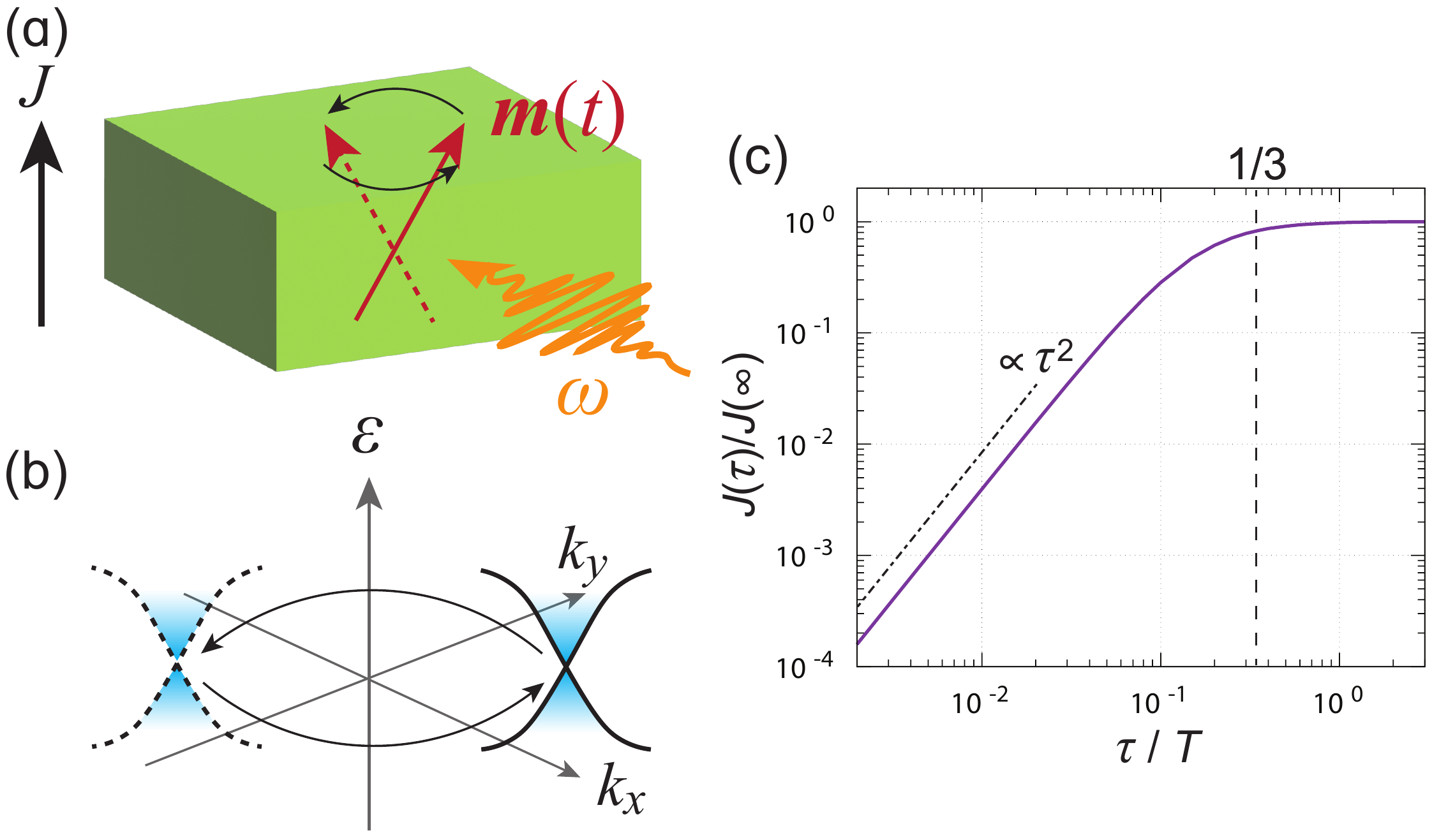}
  \caption{
  Adiabatic charge pumping and spin motive force.
  (a) Schematic of the magnetic resonance. The charge current flows along the direction of the uniform moment ($z$ axis in the main text).
  (b) Displacement of Weyl nodes by the precession of magnetic moment. The Weyl nodes shown in solid (dotted) lines corresponds to that for the magnetization in (a) shown by the arrow with solid (dotted) line. Upon magnetic precession, the Weyl node moves in the momentum space following $\bm m(t)$.
  (c) The relaxation-time dependence of the induced charge current. The result is obtained by numerically integrating Eq.~\eqref{eq:Jad}.
  The result is for $\mu/v=1$, $v_0/v=0.1$, and $J_Km_\perp/v=0.1$.
  }\label{fig:J}
\end{figure}

As a demonstration of the impact of $\bm e_{\bm kn}(t)$ induced by the magnetic dynamics, we study the Larmor precession of ferromagnetic moment in magnetic WSM.
We show that the $\bm e_{\bm kn}(t)$ field induced by the precession cause electric current $J$ along the magnetization direction.
By explicit calculation, we show $J\propto \tau^2$ dependence when the relaxation time $\tau$ is short, whereas it saturates above $\tau\sim T/3$ where $T$ is the period of the precession [Fig.~\ref{fig:J}(c)].
The saturation is a manifestation of the dissipationless current.
The low crossover $\tau/T$ implies that the long $\tau$ regime is experimentally accessible in a clean WSM if $T$ is in picoseconds.
These findings are experimentally testable in noncentrosymmetric magnetic WSM~\cite{Suzuki2019,Yang2020,Yang2021}.
The results demonstrate a unique transport phenomenon arising from topological electronic states and magnetic dynamics.

{\it Weyl Hamiltonian} --- For concreteness, we consider a tilted Weyl electron coupled to a ferromagnetic moment.
The Hamiltonian reads
\begin{align}
&H(t)=\nonumber\\
&\sum_{\bm k,\alpha,\beta} c_{\bm k\alpha}^\dagger \left[v\bm k\cdot\bm\sigma-v_0k_z\delta_{\alpha\beta}-J_K\bm m(t)\cdot\bm\sigma\right]_{\alpha\beta} c_{\bm k\beta},\label{eq:H}
\end{align}
where $\bm \sigma=(\sigma^x,\sigma^y,\sigma^z)$ ($\sigma^{x,y,z}$ are the Pauli matrices), $J_K$ is the Kondo coupling between the electrons and the ferromagnetic moment, $c_{\bm k\alpha}$ ($c_{\bm k\alpha}^\dagger$) is the annihilation (creation) operator of an electron with momentum $\bm k$, and $\bm m(t)$ [$|\bm m(t)|=1$] is the ferromagnetic moment whose direction changes over time.

The Larmor precession of the magnetic moment about $z$ axis is given by
\begin{align}
\bm m(t)=\left(m_\perp\sin(\omega t),m_\perp\cos(\omega t),\sqrt{1-m_\perp^2}\right),\label{eq:fmr}
\end{align}
where $m_\perp$ is the amplitude of precession and $\omega=2\pi/T$ is the frequency.
The steady precession of ferromagnetic moments occurs, for example, in ferromagnetic resonance as we will discuss later.

{\it Semiclassical Boltzmann theory} ---
We use the semiclassical Boltzmann theory~\cite{Xiao2010} to study the charge transport under the precession of the moment.
Within the relaxation-time approximation, the Boltzmann equation reads
\begin{align}
\partial_t f_{\bm kn}(t)=&-\frac1\tau[f_{\bm kn}(t)-f^0_{\bm kn}(t)],\label{eq:boltzmann}
\end{align}
where $\tau$ is the relaxation time, $f_{\bm kn}(t)$ is the average density of electron with momentum $\bm k$ and band index $n$ at the time $t$, and $f^0_{\bm kn}(t)=1/(e^{\beta [\varepsilon_{\bm kn}(t)-\mu(t)]}+1)$ is the Fermi distribution function for the instantaneous Hamiltonian $H(t)$.
Here, $\mu(t)$ is defined so that the electron density for instantaneous Hamiltonian $N=\int\frac{d^3k}{(2\pi)^3}f^0_{\bm kn}(t)$ conserves.

The formal solution of Eq.~\eqref{eq:boltzmann} reads
\begin{align}
f_{\bm kn}(t)=&f^0_{\bm kn}(t)e^{-\frac{t-t_0}{\tau}}+\int_{t_0}^t\frac{dt'}\tau e^{-\frac{t-t'}\tau}f^0_{\bm kn}(t'),\nonumber\\
=&\int_0^\infty\frac{dt'}\tau e^{-\frac{t'}\tau}f^0_{\bm kn}(t-t'),\label{eq:fkn}
\end{align}
where $t_0$ is the initial time at which we assume the electrons are in the equilibrium.
In the second line, we assumed $t_0/\tau\to-\infty$.
For a periodically-driven system with period $T=2\pi/\omega$, Eq.~\eqref{eq:fkn} becomes
\begin{align}
f_{\bm kn}(t)=\frac1{1-e^{-\frac{T}\tau}}\int_0^T\frac{dt'}\tau e^{-\frac{t'}\tau}f^0_{\bm kn}(t-t'),\label{eq:fkn2}
\end{align}
as $f^0_{\bm kn}(t)=f^0_{\bm kn}(t+T)$.

Within the Boltzmann theory, the electric current produced by the adiabatic pumping is given by~\cite{Xiao2010},
\begin{align}
\bm J(t)=\sum_n\int\frac{d^3k}{(2\pi)^3}q\bm e_{\bm kn}(t)f_{\bm kn}(t),
\end{align}
where $q$ is the charge of the carrier.
Combining this formula and Eq.~\eqref{eq:fkn}, the average current in a periodically-driven system reads
\begin{align}
\bar{\bm J}=&\sum_n\int\frac{d^3k}{(2\pi)^3}\int_0^T\frac{dt}{T}\frac{dt'}{\tau}\frac{qe^{-\frac{t'}\tau}}{1-e^{-\frac{T}\tau}}\bm e_{\bm kn}(t)f_{\bm kn}(t-t').\label{eq:Jad}
\end{align}
We use this formula in the rest of this paper to study the current caused by the magnetic resonance.

{\it $\tau/T\ll1$ case} ---
When $\tau/T\ll1$, Eq.~\eqref{eq:Jad} becomes
\begin{align}
\bar{\bm J}\sim&\int_0^T\frac{dt}T\int_D\frac{d^3k}{(2\pi)^3}\bm e_{\bm k+J_K\bm m(t)-\tau J_K\dot{\bm m}(t)+\tau^2 J_K\ddot{\bm m}(t),n}(t)\nonumber\\
&+{\cal O}(J_K^3).
\end{align}
Here, $D$ is the region inside the Fermi surface.
To the second order in $J_K$, the average current produced by Eq.~\eqref{eq:fmr} reads
\begin{align}
&\bar{\bm J}=\left(0,0,-\text{sgn}(v)\frac{e\tau^2\omega^3}{12\pi^2}F(v_0/v)\left(\frac{J_Km_\perp}{v}\right)^2\right),\label{eq:J1}\\
&F(x)=1+3\left(\frac1{x^2}-1\right)\left[1-\frac{\text{atanh}(x)}{x}\right].\label{eq:Fx}
\end{align}
Hence, a finite current flows by a mechanism similar to the adiabatic pump.

In a recent work, the noticeable contribution of the adiabatic pump in WSM was discussed using an analogy to the electromagnetic induction in classical electromagnetism.
When a circular electric current flows, it produces the magnetic field penetrating the circuit, known as Ampere's law.
Similarly, if a magnetic charge exists, the current of magnetic charges induces an electric field as described by an extension of Faraday's law.
This analogy also works for Weyl electrons, in which the distribution of $\bm e_{\bm kn}(t)$ field by the circular motion of Weyl nodes (a magnetic monopole of Berry curvature $\bm b_{\bm kn}$) resembles that of the magnetic field induced by the circular electric current~\cite{Ishizuka2017b}.
This argument also applies to our setup where the position of Weyl point is given by $\bm k_0=J_K\bm m(t)/v$.
However, we expect a stronger current in our setup than the previous proposal because the Kondo coupling is typically stronger than the coupling to electromagnetic fields.

We also note that the current flows along the uniform magnetization direction, and a finite tilting $v_0$ is necessary for a finite current.
The necessity of tilting is seen from the asymptotic form of $F(x)$ at $x\to0$, which is $F(x)\sim 2x^2/5$ in the $x\to0$ limit, and hence, the current is $\bar J\propto (v_0/v)^2$.
This necessity is consistent with related works, where the current flows only in the presence of warping or tilting.
However, here the current flows along the net magnetization direction, in contrast to the wave vector direction of the incident light in the previous work.
The sensitivity of electric current to the magnetization direction, rather than to the incident light, delineates the microscopic origin of the current.

We note that this current is a distinct phenomenon from the conventional spin motive force (SMF)~\cite{Stern1992,Barnes2007,Tanabe2021}.
In SMF, the electric current flows in the presence of both magnetic precession and the gradient of magnetization.
Intuitively, this phenomenon is often interpreted as a consequence fictitious electric field induced by the magnetization dynamics or the Berry phase in real space.
In contrast, the current in Eq.~\eqref{eq:J1} is a consequence of the momentum-space Berry curvature.
Reflecting the different origin, the current in Eq.~\eqref{eq:J1} does not require the magnetization gradient and flows along the direction of the net magnetization.

{\it $\tau/T\gg1$ limit} --- We next consider the long relaxation-time limit, $\tau/T\gg1$.
In this limit, the average current reads
\begin{align}
\bar{\bm J}=\int_0^T\frac{dt'dt}{T^2}\bm e_{\bm k}(t)f_{\bm kn}^0(t').
\end{align}
To the second order in $J_K$, it is
\begin{align}
\bar{\bm J}=\left(0,0,-\text{sgn}(v)\frac{e\omega}{12\pi^2}F(v_0/v)\left(\frac{J_Km_\perp}{v}\right)^2\right),\label{eq:J2}
\end{align}
where $F(x)$ is in Eq.~\eqref{eq:Fx}.
This formula is essentially equivalent to the $\tau/|t-t_0|\gg1$ case studied in a previous study~\cite{Ishizuka2017b}.
In Eq.~\ref{eq:J2}, the current is linearly proportional to $\omega$, resembling the quantization of pumped charges in the adiabatic pump.

Note that, in this limit, the current does not depend on $\tau$ as expected for the adiabatic pump.
Compared to the $\tau/T\ll1$ case, the current simply differs by a factor of $(\omega\tau)^2$.
The result implies that the current, which increases with $(\omega\tau)^2$ in the small $\tau/T$ regime, saturates as it approaches $\tau\omega=1$.
Therefore, we expect a crossover between $\tau/T\ll1$ and $\tau/T\gg1$ regimes occurs at $\tau\sim T$.

{\it Numerical calculation} ---
To further investigate the $\tau$ dependence, we evaluated the $\tau$ dependence of the current using Eq.~\eqref{eq:Jad} by numerically performing the integral.
Figure~\ref{fig:J}(c) is the result for $\mu/v=1$, $v_0/v=0.1$, and $J_Km_\perp/v=0.1$ case.
The numerical result (solid line) is proportional to $\tau^2$ in the small $\tau/T$ limit and saturates to a constant at around $\tau/T\gtrsim1/3$.
It is semi-quantitatively consistent with the above argument where the crossover occurs at $\tau\omega=2\pi\tau/T\sim1$.
In experiment, the magnetic resonance frequency ranges between $10^9$ - $10^{12}$~s$^{-1}$ while the relaxation time is in $10^{-15}$ - $10^{-12}$~s.
Hence, the adiabatic regime, $\tau/T\gg1$, might also be relevant to the experiment.

{\it Magnitude of current density} ---
We next turn to the order of current expected in magnetic WSM.
To evaluate the magnitude of current, we discuss the dynamics of the ferromagnetic moment in a ferromagnetic resonance experiment.
The dynamics of the ferromagnetic moment is phenomenologically described by Landau-Lifshits-Gilbert equation,
\begin{align}
\partial_t\bm m(t)=-\gamma\bm m(t)\times\bm B(t)+\alpha \bm m(t)\times\partial_t\bm m(t),\label{eq:LLG}
\end{align}
where $\gamma$ is the gyromagnetic constant, $\alpha$ is the Gilbert damping constant~\cite{Lenz2006,Vittoria2010}, and $\bm B(t)=(B_x(t),B_y(t),B_z)$; we assume the static magnetic field is along the $z$ axis.
Suppose the incident microwave is $B_x(t)=B_x^0\cos(t)$ and $B_y(t)=0$.
Then, at the magnetic resonance frequency $\omega=\pm\gamma B_z$, the approximate solution of Eq.~\eqref{eq:LLG} assuming $m_z(t)\sim1$ reads
\begin{align}
m_x(t)=\frac{B_x^0}{\alpha B_z}\sin(\omega t),\quad m_y(t)=\frac{B_x^0}{\alpha B_z}\cos(\omega t).
\end{align}
Hence, the amplitude of magnetic precession is $m_\perp=B_x^0/\alpha B_z$.

For the case $v=10^5$ ms$^{-1}$, $v_0=10^4$ ms$^{-1}$, $J_K=10$ meV, $B_x^0=10^{-5}$ T, $B_z=10^{-1}$ T, and $\alpha=10^{-3}$, the current density reads $J=10^{-3}$ mA cm$^{-2}$.
This estimate is several orders of magnitude larger compared to the photocurrent by the similar mechanism~\cite{Ishizuka2016,Ishizuka2017b}, despite the orders-of-magnitude smaller frequency $\omega/2\pi\sim2.8$ GHz.
Intuitively, the larger current is ascribed to the faster motion of Weyl nodes associated with the larger orbital radius.
Using the analogy to classical electromagnetism discussed above, the induced $\bm e_{\bm kn}$ field is larger for Weyl nodes moving at a higher speed.
For a fixed frequency, the Weyl nodes move faster for a larger radius of Weyl nodes, hence producing a larger $\bm e_{\bm kn}$ field that results in a larger current.

We also note that linearly-polarized electromagnetic waves can induce a finite current, in contrast to the bulk photovoltaic effects in WSM.
Recent theories for the bulk photovoltaic effect in WSM~\cite{Ishizuka2016,Chan2017,Juan2017,Bhalla2020} find that the electric current is sensitive to the polarization of the incident light. In particular, a circularly-polarized light is often necessary for a finite photocurrent.
In contrast, the above argument on magnetic precession assumes a linearly-polarized microwave; the circular motion of Weyl nodes is a consequence of the precession of magnetic moment induced by the microwave.
The result also implies that the adiabatic current discussed here is insensitive to the incident light direction, a feature distinct from the photon-drag effect.
The controllability of current by the magnetization direction and insensitiveness to the nature of incident light distinguishes this phenomenon from the photovoltaic effects generally seen in noncentrosymmetric materials~\cite{Sturman1992}.

{\it Cancellation between Weyl nodes} ---
So far, we considered the current induced by single Weyl node.
In WSM, however, multiple Weyl nodes always exist as stated in Neilsen-Ninomiya's theorem~\cite{Neilsen1981a,Neilsen1981b,Neilsen1981c}.
In addition, certain symmetries force multiple Weyl nodes to appear at the same energy canceling the electric current from each other.
For example, in a Weyl semimetal with time-reversal symmetry, two nodes with the same chirality appears at wave numbers $\bm k_0$ and $-\bm k_0$ in the momentum space.
On the other hand, in a Weyl semimetal with the inversion symmetry, two nodes with the opposite chirality appear at the two positions.
As the current in Eqs.~\eqref{eq:J1} and \eqref{eq:J2} depends on the chirality manifested in $\text{sgn}(v)$, the net current vanishes when the inversion symmetry exists.

The absence of electric current in centrosymmetric WSM is understandable from the symmetry argument of the response coefficient.
Phenomenologically, the electric current we study here is a nonlinear response to the ac magnetic field $\bar J=\sigma (B_x^0)^2$, where $\sigma$ is the nonlinear conductivity.
With the inversion operation, the current and the magnetic field transform as $\bar J\to-\bar J$ and $B_x^0\to B_x^0$, respectively.
Hence, the phenomenological formula reads $\bar J=-\sigma (B_x^0)^2$, implying $\sigma=-\sigma$, and hence, $\sigma=0$.
Therefore, a noncentrosymmetric magnetic WSM is necessary to observe the phenomenon we studied.

{\it Discussions} ---
In this work, we studied the electric current induced by the precession of magnetic moment in a magnetic Weyl semimetal.
We find that the momentum-space Berry curvature induces a finite current similar to that in the adiabatic charge pump, which is related to the circular motion of Weyl nodes in the momentum space.
The induced current is insensitive to the incident light direction and polarization, whereas it depends on the direction of the magnetic moment.
This mechanism gives rise to a finite electric current in a magnetic Weyl semimetal with a noncentrosymmetric crystal structure, where the cancellation between different Weyl-node contributions is violated by the crystal symmetry.
Finally, our estimate using a typical value for Weyl semimetal and a ferromagnet gives $\sim10^{-3}$ mA cm$^{-2}$ which should be observable in the experiments.

Experimentally, this phenomenon should be observable in a setup similar to that for spin motive force.
However, unlike the conventional spin motive force, the current due to $\bm e_{\bm kn}(t)$ appears without a magnetization gradient and the current flows along the direction of the net magnetization.
These features distinguish the $\bm e_{\bm kn}(t)$-related current from those by the conventional spin motive force.

Recent searches for Weyl semimetals discovered various materials with Weyl nodes near the Fermi level, both in noncentrosymmetric~\cite{Huang2015,Xu2015} and magnetic materials~\cite{Wan2011,Ueda2018,Shekhar2018,Belopolski2019,Liu2019,Morali2019,Suzuki2019,Yang2020,Yang2021}.
Some of these materials have both noncentrosymmetric crystal structure and magnetism, such as CeAlSi~\cite{Suzuki2019,Yang2021} and PrAlGe$_{1-x}$Si$_x$~\cite{Yang2020}.
Reflecting the absence of both inversion and time-reversal symmetries, the position of Weyl nodes in these materials are expected to be unrelated to each other~\cite{Yang2021}.
The symmetry-related cancellation is violated in these materials, allowing a larger class of phenomena related to Weyl fermions to give observable consequences.
Time-dependent Berry phase $\bm e_{\bm kn}(t)$ is one such effect that provides a route to realize nontrivial phenomena arising from the interplay of magnetism and topological electronic states.

\acknowledgements
We thank K. Burch, F. Tafti, and H. Yang for useful discussions.
This work is supported by JSPS KAKENHI (Grant Numbers JP18H03676 and JP19K14649).


\begin{thebibliography}{99}
\bibitem{He2022}
  Q. L. He, T. L. Hughes, N. P. Armitage, Y. Tokura, and K. L. Wang,
  {\it Topological spintronics and magnetoelectronics},
  Nat. Mater. {\bf21}, 15 (2022).
\bibitem{Hasan2010}
  M. Z. Hasan and C. L. Kane,
  {\it Colloquium: Topological insulators},
  Rev. Mod. Phys. 82, 3045 (2010).
\bibitem{Qi2011}
  X.-L. Qi and S.-C. Zhang,
  {\it Topological insulators and superconductors},
  Rev. Mod. Phys. {\bf83}, 1057 (2011).
\bibitem{Chang2013}
  C. Z. Chang {\it et al.},
  {\it Experimental observation of the quantum anomalous Hall effect in a magnetic topological insulator},
  Science {\bf340}, 167 (2013).
\bibitem{Checkelsky2014}
  J. G. Checkelsky {\it et al.},
  {\it Trajectory of the anomalous Hall effect towards the quantized state in a ferromagnetic topological insulator},
  Nat. Phys. {\bf10}, 731 (2014).
\bibitem{Kou2014}
  X. Kou {\it et al.},
  {\it Scale-invariant quantum anomalous Hall effect in magnetic topological insulators beyond the two-dimensional limit},
  Phys. Rev. Lett. {\bf113}, 137201 (2014).
\bibitem{Qi2008}
  X.-L. Qi, T. L. Hughes, and S.-C. Zhang,
  {\it Topological field theory of time-reversal invariant insulators},
  Phys. Rev. B {\bf78}, 195424 (2008).
\bibitem{Yan2017}
	B. Yan and C. Felser,
	Topological Materials: Weyl Semimetals
	Ann. Rev. Condens. Matter Phys. {\bf8}, 337 (2017).
\bibitem{Murakami2007}
  S. Murakami,
  {\it Phase transition between the quantum spin Hall and insulator phases in 3D: emergence of a topological gapless phase},
  New J. Phys. {\bf9}, 356 (2007).
\bibitem{Burkov2011}
  A. A. Burkov and L. Balents,
  {\it Weyl Semimetal in a Topological Insulator Multilayer},
  Phys. Rev. Lett. {\bf107}, 127205 (2011).
\bibitem{Wan2011}
  X. Wan, M. Turner, A. Vishwanath, and S. Y. Savrasov,
  {\it Topological semimetal and Fermi-arc surface states in the electronic structure of pyrochlore iridates},
  Phys. Rev. B {\bf83}, 205101 (2011).
\bibitem{Armitage2018}
	N. P. Armitage, E. J. Mele, \& A. Vishwanath,
	{\it Weyl and Dirac semimetals in three-dimensional solids},
	Rev. Mod. Phys. {\bf90}, 015001 (2018).
\bibitem{Krempa2014}
  W. Witczak-Krempa, G. Chen, Y. B. Kim, L. Balents,
  {\it Correlated quantum phenomena in the strong spin-orbit regime},
  Annu. Rev. Condens. Matter Phys. {\bf5}, 57 (2014).
\bibitem{Krempa2012}
  W. Witczak-Krempa and Y. B. Kim
  {\it Topological and magnetic phases of interacting electrons in the pyrochlore iridates},
  Phys. Rev. B {\bf85}, 045124 (2012).
\bibitem{Moon2013}
  E.-G. Moon, C. Xu, Y. B. Kim, and L. Balents,
  {\it Non-Fermi-Liquid and Topological States with Strong Spin-Orbit Coupling},
  Phys. Rev. Lett. {\bf111}, 206401 (2013).
\bibitem{Ueda2018}
  K. Ueda, R. Kaneko, J. Fujioka, H. Ishizuka, N. Nagaosa, Y. Tokura,
  {\it Spontaneous Hall effect in all-in/all-out Weyl semimetal of pyrochlore iridates},
  Nat. Commun. {\bf9}, 3032 (2018).
\bibitem{Shekhar2018}
  C. Shekhar {\it et al.},
  {\it Anomalous Hall effect in Weyl semimetal half-Heusler compounds RPtBi (R = Gd and Nd)},
  Proc. Natl. Acad. Sci. {\bf115}, 9140 (2018).
\bibitem{Takahashi2018}
  K. S. Takahashi {\it et al.},
  {\it Anomalous Hall effect derived from multiple Weyl nodes in high-mobility EuTiO$_3$ films},
  Sci. Adv. {\bf4}, eaar7880 (2018).
\bibitem{Maekawa2017}
  S. Maekawa, S. O. Valenzuela, E. Saitoh, and T. Kimura, Eds.
  {\it Spin current} 2ed.
  (Oxford Univ. Press, Oxford, 2017).
\bibitem{Araki2018}
  Y. Araki and K. Nomura,
  {\it Charge Pumping Induced by Magnetic Texture Dynamics in Weyl Semimetals},
  Phys. Rev. Appl. {\bf10}, 014007 (2018).
\bibitem{Hannukainen2020}
  J. D. Hannukainen, Y. Ferreiros, A. Cortijo, and J. H. Bardarson,
  {\it Axial anomaly generation by domain wall motion in Weyl semimetals}
  Phys. Rev. B {\bf102}, 241401(R) (2020).
\bibitem{Hannukainen2021}
  J. D. Hannukainen, A. Cortijo, J. H. Bardarson, Y. Ferreiros
  {\it Electric manipulation of domain walls in magnetic Weyl semimetals via the axial anomaly},
  SciPost Phys. {\bf10}, 102 (2021).
\bibitem{Thouless1983}
	D. J. Thouless,
	{\it Quantization of particle transport},
	Phys. Rev. B {\bf27}, 6083 (1983).
\bibitem{Niu1984}
  Q. Niu and D. J. Thouless,
  {\it Quantised adiabatic charge transport in the presence of substrate disorder and many-body interaction},
  J. Phys. A: Math. Gen. {\bf17}, 2453 (1984).
\bibitem{Sandaram1999}
	G. Sundaram and Q. Niu,
	{\it Wave-packet dynamics in slowly perturbed crystals: Gradient corrections and Berry-phase effects},
	Phys. Rev. B {\bf59}, 14915 (1999).
\bibitem{Xiao2010}
	D. Xiao, M.-C. Chang, and Q. Niu,
	{\it Berry phase effect in electronic properties},
	Rev. Mod. Phys. {\bf82}, 1959 (2010).
\bibitem{Thouless1982}
	D. J. Thouless, M. Kohmoto, M. P. Nightingale, and M. den Nijs,
	Phys. Rev. Lett. {\bf 49}, 405 (1982).
\bibitem{Karplus1954}
	R. Karplus \& J. M. Luttinger,
	{\it Hall Effect in Ferromagnetics},
	Phys. Rev. {\bf95}, 1154 (1954).
\bibitem{Son2013}
	D. T. Son \& B. Z. Spivak,
	{\it Chiral anomaly and classical negative magnetoresistance of Weyl metals},
	Phys. Rev. B {\bf88}, 104412 (2013).
\bibitem{Ishizuka2019a}
	H. Ishizuka \& N. Nagaosa,
	{\it Robustness of anomaly-related magnetoresistance in dopedWeyl semimetals},
	Phys. Rev. B {\bf99}, 115205 (2019).
\bibitem{Ishizuka2019b}
	H. Ishizuka \& N. Nagaosa,
	{\it Tilting dependence and anisotropy of anomaly-related magnetoconductance in type-II Weyl semimetals},
	Sci. Rep. {\bf9}, 16149 (2019).
\bibitem{Nakajima2016}
	S. Nakajima, T. Tomita, S. Taie, T. Ichinose, H. Ozawa, L.Wang, M. Troyer, and Y. Takahashi,
	{\it Topological Thouless pumping of ultracold fermions},
	Nat. Phys. {\bf12}, 296 (2016).
\bibitem{Lohse2016}
	M. Lohse, C. Schweizer, O. Zilberberg, M. Aidelsburger, and I. Bloch,
	{\it A Thouless quantum pump with ultracold bosonic atoms in an optical superlattice},
	Nat. Phys. {\bf12}, 350 (2016).
\bibitem{Ishizuka2016}
	H. Ishizuka, T. Hayata, M. Ueda, \& N. Nagaosa,
	{\it Emergent Electromagnetic Induction and Adiabatic Charge Pumping in Noncentrosymmetric Weyl Semimetals},
	Phys. Rev. Lett. {\bf117}, 216601 (2016).
\bibitem{Ishizuka2017b}
	H. Ishizuka, T. Hayata, M. Ueda, \& N. Nagaosa,
	{\it Momentum-space electromagnetic induction in Weyl semimetals},
	Phys. Rev. B {\bf95}, 245211 (2017).
\bibitem{Suzuki2019}
	T. Suzuki, L. Savary, J.-P. Liu, J. W. Lynn, L. Balents, \& J. G. Checkelsky,
	{\it Singular angular magnetoresistance in a magnetic nodal semimetal},
	Science {\bf365}, 377 (2019).
\bibitem{Yang2020}
    H.-Y. Yang, B. Singh, B. Lu, C.-Y. Huang, F. Bahrami, W.-C. Chiu, D. Graf, S.-M. Huang, B. Wang, H. Lin, D. Torchinsky, A. Bansil, \& F. Tafti,
	{\it Transition from intrinsic to extrinsic anomalous Hall effect in the ferromagnetic Weyl semimetal PrAlGe$_{1-x}$Si$_x$},
    APL Mater. {\bf8}, 011111 (2020).
\bibitem{Yang2021}
    H.-Y. Yang ,1 B. Singh, J. Gaudet, B. Lu , C.-Y. Huang, W.-C. Chiu, S.-M. Huang, B. Wang, F. Bahrami, B. Xu, J. Franklin, I. Sochnikov, D. E. Graf, G. Xu, Y. Zhao, C. M. Hoffman, H. Lin , D. H. Torchinsky, C. L. Broholm, A. Bansil, \& F. Tafti,
	{\it Noncollinear ferromagnetic Weyl semimetal with anisotropic anomalous Hall effect},
    Phys. Rev. B {\bf103}, 115143 (2021).
\bibitem{Stern1992}
    A. Stern,
    {\it Berry’s phase, motive forces, and mesoscopic conductivity},
    Phys. Rev. Lett. {\bf68}, 1022 (1992).
\bibitem{Barnes2007}
    S. E. Barnes and S. Maekawa,
    {\it Generalization of Faraday’s Law to Include Nonconservative Spin Forces},
    Phys. Rev. Lett. {\bf98}, 246601 (2007).
\bibitem{Tanabe2021}
	K. Tanabe \& J.-I. Ohe,
	{\it Spin-motive force in ferromagnetic and ferrimagentic materials},
	J. Phys. Soc. Jpn. {\bf90}, 081011 (2021).
\bibitem{Lenz2006}
  K. Lenz, H. Wende, W. Kuch, K. Baberschke, K. Nagy, A. J\'anossy,
  {\it Two-magnon scattering and viscous Gilbert damping in ultrathin ferromagnets},
  Phys. Rev. B {\bf 73}, 144424 (2006).
\bibitem{Vittoria2010}
  C. Vittoria, S. D. Yoon, and A. Widom,
  {\it Relaxation mechanism for ordered magnetic materials},
  Phys. Rev. B {\bf81}, 014412 (2010).
\bibitem{Chan2017}
	C.-K. Chan, N. H. Lindner, G. Refael, \& P. A. Lee,
	{\it Photocurrents in Weyl semimetals},
	Phys. Rev. B {\bf95}, 041104(R) (2017).
\bibitem{Juan2017}
	F. de Juan, A. G. Grushin, T. Morimoto \& J. E Moore,
	{\it Quantized circular photogalvanic effect in Weyl semimetals},
	Nat. Commun. {\bf8}, 15995 (2017).
\bibitem{Bhalla2020}
	P. Bhalla, A. H. MacDonald, D. Culcer,
	{\it Resonant photovoltaic effect in doped magnetic semiconductors},
	Phys. Rev. Lett. {\bf124}, 087402 (2020).
\bibitem{Sturman1992}
	B. I. Sturman, V. M. Fridkin,
	{\it The Photovoltaic and Photorefractive Effects in Noncentrosymmetric Materials},
	(Gordon and Breach Science Publishers, Amsterdam, Netherlands, 1992).
\bibitem{Neilsen1981a}
  H. B. Nielsen and M. Ninomiya,
  {\it Absence of Neutrinos on a Lattice. 1. Proof by Homotopy Theory},
  Nucl. Phys. B {\bf185}, 20 (1981).
\bibitem{Neilsen1981b}
  H. B. Nielsen and M. Ninomiya,
  {\it No Go Theorem for Regularizing Chiral Fermions},
  Phys. Lett. {\bf105B}, 219 (1981).
\bibitem{Neilsen1981c}
  H. B. Nielsen and M. Ninomiya,
  {\it Absence of Neutrinos on a Lattice. 2. Intuitive Topological Proof},
  Nucl. Phys. {\bf B193}, 173 (1981).
\bibitem{Huang2015}
  S.-M. Huang, S.-Y. Xu, I. Belopolski, C.-C. Lee, G. Chang, B. Wang, N. Alidoust, G. Bian, M. Neupane, C. Zhang, S. Jia, A. Bansil, H. Lin, and M. Zahid Hasan,
  {\it A Weyl Fermion semimetal with surface Fermi arcs in the transition metal monopnictide TaAs class},
  Nat. Commun. {\bf6}, 7373 (2015).
\bibitem{Xu2015}
  S.-Y. Xu, I. Belopolski, N. Alidoust, M. Neupane, G. Bian, C. Zhang, R. Sankar, G. Chang, Z. Yuan, C. C. Lee, S.-M. Huang, H. Zheng, J. Ma, D. S. Sanchez, B.Wang, A. Bansil, F. Chou, P. P. Shibayev, H. Lin, S. Jia, and M. Zahid Hasan,
  {\it Discovery of a Weyl fermion semimetal and topological Fermi arcs},
  Science {\bf349}, 613 (2015).
\bibitem{Belopolski2019}
  I. Belopolski, K. Manna, D. S. Sanchez, G. Chang, B. Ernst, J. Yin, S. S. Zhang, T. Cochran, N. Shumiya, H. Zheng, B. Singh, G. Bian, D. Multer, M. Litskevich, X. Zhou, S.-M. Huang, B. Wang, T.-R. Chang, S.-Y. Xu, A. Bansil, C. Felser, H. Lin, and M. Z. Hasan,
  {\it Discovery of topological Weyl fermion lines and drumhead surface states in a room temperature magnet},
  Science {\bf365}, 1278 (2019).
\bibitem{Liu2019}
  D. F. Liu, A. J. Liang, E. K. Liu, Q. N. Xu, Y. W. Li, C. Chen, D. Pei, W. J. Shi, S. K. Mo, P. Dudin, T. Kim, C. Cacho, G. Li, Y. Sun, L. X. Yang, Z. K. Liu, S. S. P. Parkin, C. Felser, Y. L. Chen,
  {\it Magnetic Weyl semimetal phase in a Kagom\'e crystal},
  Science {\bf365}, 1282 (2019).
\bibitem{Morali2019}
  N. Morali, R. Batabyal, P. K. Nag, E. Liu, Q. Xu, Y. Sun, B. Yan, C. Felser, N. Avraham, H. Beidenkopf,
  {\it Fermi-arc diversity on surface terminations of the magnetic Weyl semimetal Co$_3$Sn$_2$S$_2$},
  Science {\bf365}, 1286 (2019).


\end{thebibliography}
\end{document}